\newif\ifss
\newif\ifblinded
\newif\ifaoas

\sstrue
\blindedfalse

\ifss
  \documentclass{article}
\else \ifaoas
  \documentclass[aoas]{imsart}
\else
  \documentclass[12pt, letterpaper]{article}
  \usepackage[margin = 1in]{geometry}
\fi
\fi

\usepackage{amsmath,bm}
\usepackage{amsthm}
\usepackage{graphicx}
\usepackage{amsfonts}
\usepackage{amssymb}
\usepackage{latexsym}
\usepackage[usenames,dvipsnames]{color}
\usepackage{mathrsfs}
\usepackage[authoryear]{natbib}
\usepackage{lineno}
\usepackage{tikz}
\usepackage{algorithm} \usepackage{algorithmicx}
\usepackage{algpseudocode}
\usepackage{versions}
\usepackage{xcolor}
\usepackage{dsfont}
\usepackage{titling}
\usepackage{fancyhdr}
\usepackage{multibib}
\newcites{New}{References}
\ifaoas \else
\usepackage{setspace}
\fi
\usepackage[shortlabels]{enumitem}
\usepackage[T1]{fontenc}
\usepackage{ifthen}
\usepackage{xcomment}
\usepackage{blindtext}
\usepackage{microtype}
\usepackage{booktabs}
\usepackage[colorinlistoftodos]{todonotes}
\usepackage{soul}
\usepackage{pdfpages}

\newcommand*\patchAmsMathEnvironmentForLineno[1]{%
  \expandafter\let\csname old#1\expandafter\endcsname\csname #1\endcsname
  \expandafter\let\csname oldend#1\expandafter\endcsname\csname end#1\endcsname
  \renewenvironment{#1}%
     {\linenomath\csname old#1\endcsname}%
     {\csname oldend#1\endcsname\endlinenomath}}%
\newcommand*\patchBothAmsMathEnvironmentsForLineno[1]{%
  \patchAmsMathEnvironmentForLineno{#1}%
  \patchAmsMathEnvironmentForLineno{#1*}}%
\AtBeginDocument{%
\patchBothAmsMathEnvironmentsForLineno{equation}%
\patchBothAmsMathEnvironmentsForLineno{align}%
\patchBothAmsMathEnvironmentsForLineno{flalign}%
\patchBothAmsMathEnvironmentsForLineno{alignat}%
\patchBothAmsMathEnvironmentsForLineno{gather}%
\patchBothAmsMathEnvironmentsForLineno{multline}%
}

\usepackage[authoryear]{natbib}

\definecolor{pal1}{cmyk}{.90,.30,.0,.0}
\definecolor{pal2}{cmyk}{0.80, 0.00, 1.00, 0.00}
\definecolor{pal3}{cmyk}{0.00, 0.40, 0.25, 0.00}
\definecolor{pal4}{cmyk}{0.10,0.90,0.80,0.00}
\definecolor{palg}{HTML}{97F99F}
\definecolor{palb}{HTML}{B3F0FF}
\definecolor{palorange}{HTML}{FFC36B}

\newcommand{\Categorical}{\operatorname{Categorical}}

\newcommand{\Normal}{\operatorname{Gau}}
\newcommand{\Gam}{\operatorname{Ga}}

\newcommand{\Dirichlet}{\operatorname{Dirichlet}}

\newcommand{\TruncatedNormal}{\operatorname{TruncGau}}

\newcommand{\Var}{\operatorname{Var}}
\newcommand{\Cov}{\operatorname{Cov}}

\newcommand{\diag}{\operatorname{diag}}

\newcommand{\trans}{^{\top}}

\newcommand{\supp}[1]{^{(#1)}}

\newcommand{\iid}{\stackrel{\textnormal{iid}}{\sim}}
\newcommand{\indep}{\stackrel{\textnormal{indep}}{\sim}}

\newcommand{\Identity}{\operatorname{I}}
\newcommand{\ones}{\mathds 1}
\newcommand{\zeros}{\bm 0}

\newcommand{\Reals}{\mathbb R}

\theoremstyle{definition}

\usepackage{hyperref}
\hypersetup{colorlinks=true,allcolors=blue}
\usepackage{hypcap}

\renewcommand{\vec}{\operatorname{vec}}

\newcommand{\sS}{\mathcal S}
\newcommand{\sI}{\mathcal I}
\newcommand{\sU}{\mathcal U}

\newcommand{\Yelp}{\texttt{Yelp!}}
\newcommand{\Data}{\mathcal D}

\ifaoas
\else \ifblinded
\else
  \author{
    Antonio R. Linero\thanks{
      Department of Statistics, Florida State University
    } \thanks{Email: \texttt{arlinero@stat.fsu.edu}}\ ,
    Jonathan R. Bradley$^{*}$,
    and
    Apurva Desai$^{*}$
  }
\fi
\fi

\ifaoas
\else
  \title{
    \vspace{-3em}
    Multi-rubric Models for Ordinal Spatial Data with Application to Online Ratings Data
  }
\fi

\begin{document}



\ifaoas

  \begin{frontmatter}
  \title{Multi-rubric Models for Ordinal Spatial Data with Application to Online Ratings Data}
  \runtitle{Multi-rubric Models}
  \author{\fnms{Antonio R.} \snm{Linero}\ead[label=e1]{arlinero@stat.fsu.edu}},
  \author{\fnms{Jonathan R.} \snm{Bradley}}
  \and
  \author{\fnms{Apurva} \snm{Desai}}
  \runauthor{Linero et al.}
  \affiliation{Florida State University}

\else
  \maketitle
\fi 
\begin{abstract}
  \noindent
  Interest in online rating data has increased in recent years in which ordinal ratings of products or local businesses are provided by users of a website, such as \Yelp\ or \texttt{Amazon}. One source of heterogeneity in ratings is that users apply different standards when supplying their ratings; even if two users benefit from a product the same amount, they may translate their benefit into ratings in different ways. In this article we propose an ordinal data model, which we refer to as a multi-rubric model, which treats the criteria used to convert a latent utility into a rating as user-specific random effects, with the distribution of these random effects being modeled nonparametrically. We demonstrate that this approach is capable of accounting for this type of variability in addition to usual sources of heterogeneity due to item quality, user biases, interactions between items and users, and the spatial structure of the users and items. We apply the model developed here to publicly available data from the website \Yelp\ and demonstrate that it produces interpretable clusterings of users according to their rating behavior, in addition to providing better predictions of ratings and better summaries of overall item quality.

  \ifaoas \else
    \bigskip

    

  \fi
\end{abstract}

\ifaoas
\end{frontmatter}
\else
\ifss
\else
\doublespacing
\fi
\fi

\section{Introduction}
\label{sec:introduction}

In recent years, the complexity of data used to make decisions has increased dramatically. A prime example of this is the use of online reviews to decide whether to purchase a product or visit a local business; we refer to the objects being reviewed as \emph{items}. Consider data provided by \Yelp\ (see, \url{http://www.yelp.com/}), which allows users to rate items, such as restaurants, convenience stores, and so forth, on a discrete scale from one to five ``stars.'' Additional features of the businesses are also known, such as the spatial location and type of business.
Datasets of this type are typically very large and exhibit complex dependencies.

As an example of this complexity, users of \Yelp\ effectively determine their own standards when rating a local business. We refer to the particular standards a user applies as a \emph{rubric}. We might imagine a latent variable \(Y_{iu}\) representing the \emph{utility}, or benefit, user \(u\) obtained from item \(i\). For a given level of utility, however, different users may still give different ratings due to having different standards for the ratings; for example, one user may rate a restaurant 5 stars as long as it provides a non-offensive experience, a second user might require an exceptional experience to rate the same restaurant 5 stars, and a third user may rate all items with \(1\) star in order to ``troll'' the website. Each of these users are applying different rubrics in translating their utility to a rating for the restaurant. In addition we also expect user-specific selection bias in the sense that some users may rate every restaurant they attend, while other users may only rate restaurants that they feel strongly about.

This article makes several contributions. First, we develop a semiparametric Bayesian model which accounts for the existence of multiple rubrics for ratings data that are observed over multiple locations. To do this, we use a spatial cumulative probit model \citep[e.g., see][]{Higgs, BerretProbit,schliep2015data} in which the break-points are modeled as user-specific random effects. This requires a flexible model for the distribution \(F\) of the random effects, which we model as a discrete mixture.  A by-product of our approach is that we obtain a clustering of users according to the rubrics they are using.

Second, we use the multi-rubric model to address novel inferential questions. For example, ratings provided to a user might be adjusted to match that user's rubric, or to provide a distribution for the rating that a user would provide conditional on having a particular rubric. Utilizing this user-specific standardization of ratings may provide users with better intuition for the overall quality of an item.

This adjustment of restaurant quality for the rubrics is similar to, but distinct from, the task of predicting a user's ratings. Good predictive performance is required for \emph{filtering}, which refers to the task of processing the rating history of a user and producing a list of recommendations \citep[for a review, see][]{bobadilla2013recommender}. As a third contribution, we show that allowing for multiple rubrics improves predictions.

The model proposed here also has interesting statistical features. A useful feature of our model is that it allows for more accurate comparisons across items. For example, if a user rates all items with \(1\) star, then the model discounts this user's ratings. This behavior is desirable for two reasons. First, if a user genuinely rates all items with \(1\) star, then their rating is unhelpful. Second, it down-weights the ratings of users who are exhibiting selection bias and only rating items which they feel strongly about, which is desirable as comparisons across items will be more indicative of true quality if they are based on individuals who are not exhibiting large degrees of selection bias. 

Additionally, the rubrics themselves may be of intrinsic interest. We demonstrate that the rubrics learned by our model are highly interpretable. For example, when analyzing the \Yelp\ dataset in Section~\ref{sec:data-analysis}, we obtain Figure~\ref{fig:rubric-props} which displays the ratings observed for users assigned to a discrete collection of rubrics and reveals several distinct rating patterns displayed by users.

Other features of our model are also of potentially independent interest. The multi-rubric model can be interpreted as a novel semiparametric random-effects model for ordinal data, even for problems in which the intuition behind the multi-rubric model in terms of latent utility does not hold. Other study designs in which the multi-rubric analogy may be useful include longitudinal survey studies, or more general ordinal repeated-measures designs. Additionally, the cumulative probit model we use to model latent user preferences includes a spatial process to account for spatial dependencies across local businesses. Recovering an underlying spatial process allows for recommending entire regions to visit, rather than singular items. The development of low-rank spatial methodology for large-scale dependent ordinal data is of interest within the spatial literature, as the current spatial literature for ordinal data do not typically address large datasets on a similar order of the \Yelp\ dataset \citep[e.g., see][among others]{de2004simple, de2000bayesian, Chen2000, Cargnoni, KnorrHeld, CarlinDiscreteCat, Higgs, BerretProbit, rainfallCat}. We model the underlying spatial process using a low-rank approximation \citep{johan} to a desired Gaussian process \citep{banerjee,bradleyMSTM}.

Starting from \citet{koren2011ordrec}, several works in the recommender systems literature have considered ordinal matrix factorization (OMF) procedures which are similar in many respects to our model (see also \citealt*{paquet2012hierarchical} and \citealt*{houlsby2014cold}). Our work differs non-trivially from these works in that the multi-rubric model treats the break-points as user-specific random effects, with a nonparametric prior used for the random effects distribution \(F\). For the \Yelp\ dataset, this extra flexibility leads to improved predictive performance. Additionally, our focus in this work extends to inferential goals beyond prediction; for example, depending on the distribution of the rubrics of users who rate a given item, the estimate of overall quality for that item can be shrunk to a variety of different centers, producing novel multiple-shrinkage effects. Several works in the Bayesian nonparametric literature have also considered flexible models for random effects in multivariate ordinal models \citep{kottas2005, deyoreo2014bayesian, bao2015bayesian}, but do not treat the break-points themselves as random effects.


The paper is organized as follows. In Section~\ref{sec:multi-rubric}, we develop the multi-rubric model, with an eye towards the \Yelp\ dataset, and provide implementation details. In Section~\ref{sec:simulation}, we illustrate the methodology on synthetic data designed to mirror features of the \Yelp\ dataset, and demonstrate that we can accurately recover the number and structure of the rubrics when the model holds, as well as effectively estimate the underlying latent utility field. In Section \ref{sec:data-analysis}, we illustrate the methodology on the \Yelp\ dataset. We conclude with a discussion in Section~\ref{sec:discussion}. In supplementary material, we present simulation experiments which demonstrate identifiability of key components of the model.

\section{The Multi-rubric model}
\label{sec:multi-rubric}

\subsection{Preliminary notation}

We consider ordinal response variables \(Z_{iu}\) taking values in \(\{1, \ldots, K\}\). In the context of online ratings data, \(Z_{iu}\) represents the rating that user \(u\) provides for item \(i\). In the context of survey data, on the other hand, \(Z_{iu}\) might represent the response subject \(u\) gives to question \(i\). We do not assume that \(Z_{iu}\) is observed for all \((i,u)\) pairs, but instead we observe \((i,u) \in \mathcal S \subseteq \{1, \ldots, I\} \times \{1, \ldots, U\}\), where \(U\) is the total number of subjects and \(I\) is the total number of items. For fixed \(i\) we let \(\sU_i = \{u : (i,u) \in \sS\}\) be the set of users that rate item \(i\), and similarly for fixed \(u\) we let \(\sI_u = \{i: (i,u) \in \sS\}\) be the set of items that user \(u\) rates.

\subsection{Review of Cumulative Probit Models}
\label{sec:review-probit}

Cumulative probit models \citep{chib-93,albert1997bayesian} provide a convenient framework for modeling ordinal rating data. Consider the univariate setting, with ordinal observations \(\{Z_i : 1 \le i \le N\}\) taking values in \(\{1,\ldots,K\}\). We assume that \(Z_i\) is a rounded version of a latent variable \(Y_i\) such that \(Z_i = k\) if \(\theta_{k-1} \le Y_i < \theta_k\). Here, \(-\infty = \theta_0 \le \theta_1 \le \cdots \le \theta_K = \infty\) are unknown break-points. When \(Y_i\) has the Gaussian distribution \(Y_i \sim \Normal(x_i\trans\gamma, 1)\) this leads to the ordinal probit model, where \(\Pr(Z_i = k \mid \theta, \gamma) = \Phi(\theta_k - x_i\trans\gamma) - \Phi(\theta_{k-1} - x_i\trans\gamma)\).

We assume $\Var(Y_i) = 1$, as the variance of \(Y_i\) is confounded with the break-points \(\theta = (\theta_1, \ldots, \theta_{K-1})\). Any global intercept term is also confounded with the \(\theta\)'s; there are two resolutions to this issue. The first is to fix one of the \(\theta_k\)'s, e.g., \(\theta_1 \equiv 0\). The second is to exclude an intercept term from \(x_i\). While the former approach is often taken \citep{albert1997bayesian, Higgs}, it is more convenient in the multi-rubric setting to use the latter approach to avoid placing asymmetric restrictions on the break-points.

The ordinal probit model is convenient for Bayesian inference in part because it admits a simple data augmentation algorithm which iterates between sampling \(Y_i \indep \TruncatedNormal(x_i\trans\gamma, 1, \theta_{Z_i-1}, \theta_{Z_i})\) for \(1 \le i \le N\) and, assuming a flat prior for \(\gamma\), sampling
\begin{math}
  \gamma
  \sim
  \Normal\{(X\trans X)^{-1}X\trans \bm Y, (X\trans X)^{-1}\},
\end{math}
where \(X\) has \(i^{\text{th}}\) row \(x_i\trans\) and \(\bm Y = (Y_1, \ldots,
Y_N)\). Here, \(\TruncatedNormal(\mu, \sigma^2, a, b)\) denotes the Gaussian
distribution truncated to the interval \((a,b)\). Additionally, an update for
\(\theta\) is
needed. 
Efficient updates for \(\theta\) can be implemented by using a Metropolis-within-Gibbs step to update \(\theta\) as a block (for details, see \citealp{albert1997bayesian}, as well as \citealp{cowles1996accelerating} for alternative MCMC schemes).

\subsection{Description of the proposed model}
\label{sec:description}

\subsubsection{The multi-rubric model}
We develop an extension of the cumulative probit model to generic repeated-measures ordinal data \(\{Z_{iu} : (i,u) \in \mathcal S\}\). Following \citet{albert1997bayesian} we introduce latent utilities \(Y_{iu}\), but specify a generic ANOVA model
\begin{align}
  \label{eq:anova}
  Y_{iu} = f_{iu} + \nu_u + \xi_i + \epsilon_{iu}, \qquad \epsilon_{iu} \iid \Normal(0, 1),
\end{align}
where \(\nu_u\) and \(\xi_i\) are main effects and \(f_{iu}\) is an interaction effect. The multi-rubric model modifies the cumulative probit model by replacing the break-point parameter \(\theta\) with \(u\)-specific random effects \(\theta_u = (\theta_{u0}, \ldots, \theta_{uK})\) with \([\theta_u \mid F] \indep F\) for some unknown \(F\). As before, we let \(Z_{iu} = k\) if \(\theta_{u(k-1)} \le Y_{iu} \le \theta_{uk}\). 

For concreteness, we take \(F\) to be a finite mixture \(F = \sum_{m = 1}^M \omega_m \delta_{\theta^{(m)}}\) for some large \(M\), with \(\theta^{(m)} \iid H\) and \(\omega \sim \Dirichlet(a, \ldots, a)\), where \(\delta_{\theta\supp m}\) is a point-mass distribution at \(\theta\supp m\). We note that it is also straight-forward to use a nonparametric prior for \(F\) such as a Dirichlet process \citep{escobarwest1995, ferguson1973}. We refer to the random effects \(\theta^{(1)}, \ldots, \theta^{(M)}\) as \emph{rubrics}. Note that for each subject \(u\) there exists a latent class \(m\) such that \(\theta_u = \theta^{(m)}\).

\begin{figure}
  \centering
  \includegraphics[width=.8\textwidth]{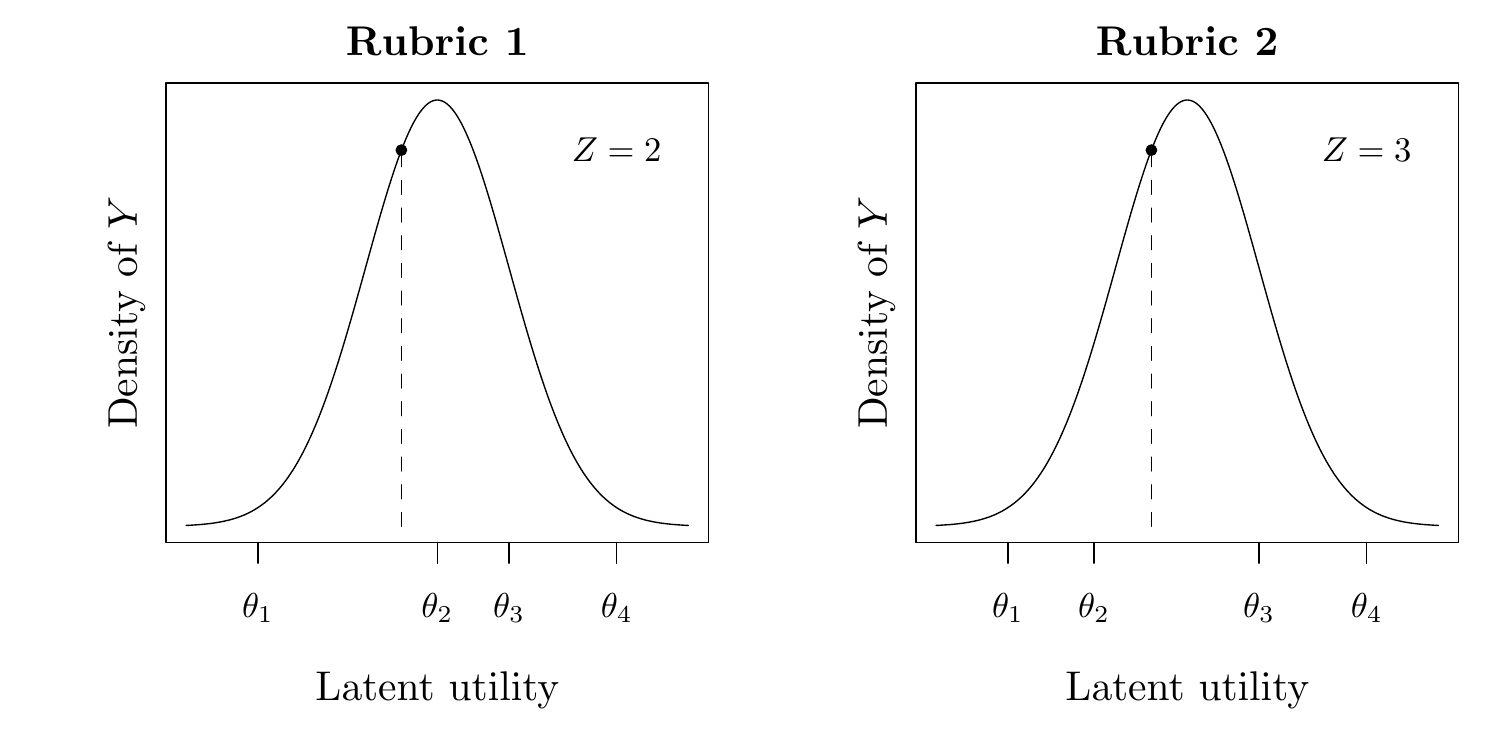}
  \caption{Visualization of the multi-rubric model. The point on the density
    indicates the realized value of \(Y\).}
  \label{fig:multi-rubric-illustration}
\end{figure}

Figure~\ref{fig:multi-rubric-illustration} displays the essential idea for the model. Viewing \(Y\) as a latent utility, the rubric that the user is associated to leads to different values of the observed rating \(Z\). In this example, the second rubric is associated to users who rate many items with a \(3\), while the first rubric is associated to users who do not rate many items with a \(3\).

Treating the break-points as random effects has several benefits. First, it offers additional flexibility over approaches for ordinal data which incorporate a random intercept \citep{gill2009nonparametric}. Due to the fact that the \(\theta_u\)'s are confounded with both the location and scale of \(Y_{iu}\), treating the break-points as random effects is at least as flexible as treating the location and scale of the distribution of the \(Y_{iu}\)'s as random effects. We require this additional flexibility, as merely treating the location and scale of the \(Y_{iu}\)'s as random effects does not allow for the variety of rating behaviors exhibited by users. By treating the break-points as random effects, we are able to capture \emph{any} distribution of ratings in a given rubric (see, e.g., Figure~\ref{fig:rubric-props}). In addition to flexibility, specifying \(F\) as a discrete mixture induces a clustering of users into latent classes. To each user \(u\) we associate a latent variable \(C_u\) such that \(C_u = m\) if \(\theta_u = \theta^{(m)}\). As will be demonstrated in Section~\ref{sec:data-analysis}, the latent classes of users discovered in this way are highly interpretable.

\subsubsection{Model for the \Yelp\ data}

Our model for the \Yelp\ data takes \(Y_{iu} \sim \Normal(\mu_{iu}, 1)\) where
\begin{align*}
  \mu_{iu} =
  x_i\trans\gamma +
  \alpha_u\trans\beta_i +
  W_i +
  b_i,
  \qquad
  W_i = \psi(s_i)\trans\eta.
\end{align*}
This is model \eqref{eq:anova} with \(f_{iu} = \alpha_u\trans\beta_i\), \(\xi_i = x_i\trans\gamma + W_i + b_i\), and \(\nu_u\) removed. This model can be motivated as a combination of the fixed-rank kriging approach of \citet{johan} with the probabilistic matrix factorization approach of \citet{salakhutdinov2007probabilistic}. The terms \(x_i\trans\gamma\), \(W_i\), and \(b_i\) are used to account for heterogeneity in the items. The term \(x_i\trans\gamma\) accounts for known covariates \(x_i \in \Reals^p\) associated to each item. The term \(W_i\) is used to capture spatial structure, and is modeled with a basis function expansion \(W_i = \psi(s_i)\trans\eta\) where \(s_i\) denotes the longitude-latitude coordinates associated to the item and \(\psi(s) = (\psi_1(s) \ldots, \psi_r(s))\trans\) is a vector of basis functions. We note that it is straight-forward to replace our low-rank approach for \(W_i\) with more elaborate approaches such as the full-scale approach of \citet{sang2012full}. The term \(b_i\) is an item-specific random effect which is used to capture item heterogeneity which cannot be accounted for by the covariates or the low-rank spatial structure.

The vectors \(\alpha_u\) and \(\beta_i\) intuitively correspond to unmeasured user-specific and item-specific latent features. The term \(\alpha_u\trans\beta_i\) is large/positive when \(\alpha_u\) and \(\beta_i\) point in the same direction (i.e., the user's preferences align with the item's characteristics), and is large/negative when \(\alpha_u\) and \(\beta_i\) point in opposite directions. This allows the model to account not only for user-specific biases (\(\theta_u\)) and item-specific biases \((x_i, W_i, b_i)\), but also interaction effects.

The multi-rubric model can be summarized by the following hierarchical model. For each model, we implicitly assume the statements hold conditionally on all variables in the models below, and that conditional independence holds within each model unless otherwise stated.
\begin{description}
\item[Response model:] \(Z_{iu} = k\) with probability \(w_{iuk} = \Phi(\theta_{uk} - \mu_{iu}) - \Phi(\theta_{u(k-1)} - \mu_{iu})\) and \(\mu_{iu} = x_i\trans\gamma + \alpha_u\trans\beta_i + W_i + b_i\).
\item[Random effect model:] \(\theta_u \iid F\), \(\alpha_{u} \sim \Normal(0,\sigma^2_\alpha \Identity)\), \(\beta_{i} \sim \Normal(0,\sigma^2_\beta \Identity)\), and \(b_i \sim \Normal(0, \sigma^2_b)\).
\item[Spatial process model:] \(W_i = \psi(s_i)\trans \eta\) where \(\eta \sim \Normal(0, \Sigma_\eta)\).
\item[Parameter model:] \(\gamma \sim \operatorname{Flat}\) and \(F = \sum_{m=1}^M \omega_m \delta_{\theta^{(m)}}\) where \(\omega \sim \Dirichlet(a,\ldots,a)\) and \(\theta^{(m)} \iid H\).
\end{description}

To complete the model we must specify values for the hyperparameters \(\sigma_\alpha, \sigma_\beta, \sigma_b, \Sigma_\eta, a\), and \(H\), as well as the number of rubrics \(M\) and the number of latent factors \(L\). In our illustrations we place half-Gaussian priors for the scale parameters, with \((\sigma_\beta, \sigma_b) \iid \Normal_+(0, 1)\), and \(\sigma_\alpha \equiv 1\). We let \(\Sigma_\eta = \diag(\sigma^2_\eta, \ldots, \sigma^2_\eta)\) and set \(\sigma_\eta \sim \Normal_+(0,1)\). Here, \(\Normal_+(0,1)\) denotes a standard Gaussian distribution truncated to the positive reals. For a discussion of prior specification for variance parameters, see \citet{gelman2006prior} and \citet{simpson2017penalising}.

In our illustrations we use \(M = 20\). For the \Yelp\ dataset, the choice of \(M = 20\) rubrics is conservative, and by setting \(a = \kappa / M\) for some fixed \(\kappa > 0\), we encourage \(\omega\) to be nearly-sparse \citep{ishwaran2002dirichlet, linero2016bayesian}. This strategy effectively lets the data determine how many rubrics are needed, as the prior encourages \(\omega_m \approx 0\) if rubric \(m\) is not needed. The prior \(H\) for \(\theta^{(1)}, \ldots, \theta^{(M)}\) is chosen to have density \(h(\theta) = \prod_{k=1}^K \Normal(\theta_k \mid 0, \sigma_\theta^2) I(\theta_1 \le \cdots \le \theta_{K-1})\) so that \(\theta^{(m)}\) has the distribution of the order statistics of \(K-1\) independent \(\Normal(0, \sigma^2_\theta)\) variables.

\subsection{Evaluating item quality}

A commonly used measure of item quality is the average rating of a user from the population \(\lambda_i = E(Z_{iu} \mid x_i, \phi_i, \gamma)\) where \(\phi_i = (\beta_i, b_i, W_i)\). This quantity is given by
\begin{align*}
  \lambda_i 
  &= \sum_{k = 1}^K k \cdot \Pr(Z_{iu} = k \mid x_i, \phi_i, \gamma)
  \\&= \sum_{k=1}^K 
      \sum_{m=1}^M k \cdot \omega_m \cdot 
      \int \Pr(Z_{iu} = k \mid x_i, \phi_i, \alpha_u, \gamma, C_u = m)
      \, \Normal(\alpha \mid 0, \sigma^2_\alpha \Identity) \ d\alpha.
\end{align*}
Using properties of the Gaussian distribution, and recalling that \(\sigma^2_\alpha = 1\), it can be shown that
\begin{align}
  \label{eq:lambda}
  \lambda_i 
  = 
  \sum_{k=1}^K \sum_{m = 1}^M k \cdot \omega_m \cdot 
  \left\lbrace 
  \Phi\left(\frac{\theta^{(m)}_k - \xi_i}{\sqrt{1 + \|\beta_i\|^2}}\right) - 
    \Phi\left(\frac{\theta^{(m)}_{k-1} - \xi_i}{\sqrt{1 + \|\beta_i\|^2}}\right)
  \right\rbrace,
\end{align}
where \(\xi_i = x_i\trans\gamma + b_i + W_i\). In Section~\ref{sec:data-analysis}, we demonstrate the particular users who rated item \(i\) exert a strong influence on the \(\lambda_i\)'s, particularly for restaurants with few ratings.

Rather than focusing on an omnibus measure of overall quality, we can also adjust the overall quality of an item to be rubric-specific. This amounts to calculating
\begin{math}
  \lambda_{im} = E(Z_{iu} \mid x_i, \phi_i, \gamma, C_u = m), 
\end{math}
which represents the average rating of item \(i\) if all used rubric \(m\). Similar to \eqref{eq:lambda}, this quantity can be computed as
\begin{align}
  \label{eq:lambda-k}
  \lambda_{im}
  = 
  \sum_{k=1}^K  k \cdot 
  \left\lbrace 
  \Phi\left(\frac{\theta^{(m)}_k - \xi_i}{\sqrt{1 + \|\beta_i\|^2}}\right) - 
    \Phi\left(\frac{\theta^{(m)}_{k-1} - \xi_i}{\sqrt{1 + \|\beta_i\|^2}}\right)
  \right\rbrace.
\end{align}
In Section~\ref{sec:data-analysis}, we use both \eqref{eq:lambda} and \eqref{eq:lambda-k} to understand the statistical features of the multi-rubric model.

\subsection{Implementation Details}
We use the reduced rank model \(W = \Psi\eta + b\) where \(\Psi\in\Reals^{I \times r}\) has \(i^{\text{th}}\) row given by \(\psi(s_i)\trans\). We choose \(\Psi\) so that \(\Cov(\Psi\eta)\) is an optimal low-rank approximation to \(\sigma^2_\eta \Xi\) where \(\Xi\) is associated to a target positive semi-definite covariogram. This is accomplished by taking \(\Psi\) composed of the first \(r\) columns of \(\Gamma D^{1/2}\) where \(\Xi = \Gamma D \Gamma\trans\) is the spectral decomposition of \(\Xi\). The Eckart-Young-Mirsky theorem states that this approximation is optimal with respect to both the operator norm and Frobenius norm \citep[see, e.g.,][Chapter 8]{rasmussen2005gaussian}. A similar strategy is used by \citet{bradleyPCOS,bradleyMSTM}, who use an optimal low-rank approximation of a target covariance structure \(\Xi \approx \Psi \Sigma_\eta \Psi\trans\) where the basis \(\Psi\) is held fixed but \(\Sigma_{\eta}\) is allowed to vary over all positive-definite \(r \times r\) matrices. In our illustrations, we use the squared-exponential covariance, i.e., $\Xi_{ij} = \exp({-{\rho}\|s_i - s_j\|^2})$ \citep{cressie2015statistics}.

To complete the specification of the model, we must specify the bandwidth \(\rho\), the number of latent factors \(L\), and the number of basis functions \(r\). We regard \(L\) as a tuning parameter, which can be selected by assessing prediction performance on a held-out subset of the data. In principle, a prior can be placed on \(\rho\), however this results in a large computational burden; we instead evaluate several fixed values of \(\rho\) chosen according to some rules-of-thumb and select the value with the best performance. For the \Yelp\ dataset, we selected \(\rho = 1000\), which corresponds undersmoothing the spatial field relative to Scott's rule \citep[see, e.g.,][]{hardle2000multivariate} by roughly a factor of two, and remark that substantively similar results are obtained with other bandwidths. Finally, \(r\) can be selected so that the proportion of the variance \(\sum_{d=1}^r D_{ii}^2 / \sum_{d=1}^n D_{ii}^2\) in $\Xi$ accounted for by the low-rank approximation exceeds some preset threshold; for the \Yelp\ dataset, we chose \(r = 500\) to account for 99\% of the variance in \(\Xi\).

When specifying the number of rubrics \(M\), we have found that the model is most reliable when \(M\) is chosen large and \(a = \kappa / M\) for some \(\kappa > 0\); under these conditions, the prior for \(F\) is approximately a Dirichlet process with concentration \(\kappa\) and base measure \(H\) (see, e.g., \citealp{teh2006hierarchical}). We recommend choosing \(M\) to be conservatively large and allowing the model to remove unneeded rubrics through the sparsity-inducing prior on \(\omega\). We have found that taking \(M\) large is necessary for good performance even in simulations in which the true number of rubrics is small and known.


We use Markov chain Monte Carlo to approximately sample from the posterior distribution of the parameters. A description of the sampler is given in the appendix.

\subsection{A note on selection bias}

Let \(\Delta_{iu} =1\) if \((i,u) \in \sS\), and \(\Delta_{iu} = 0\) otherwise. In not modeling the distribution of \(\Delta_{iu}\), we are implicitly modeling the distribution of the \(Z_{iu}\)'s conditional on \(\Delta_{iu} = 1\). When selection bias is present, this may be quite different than the marginal distribution of \(Z_{iu}\)'s. Experiments due to \citet{marlin2009collaborative} provide evidence that selection bias may be present in practice.

A useful feature of the approach presented here is that it naturally down-weights users who are exhibiting selection bias. For example, if user \(u\) only rates items they feel negatively about, they will be assigned to a rubric \(m\) for which \(\theta^{(m)}_{1}\) is very large; this has the effect of ignoring their ratings, as there will be effectively no information in the data about their latent utility. As a result, when estimating overall item quality, the model naturally filters out users who are exhibiting extreme selection bias, which may be desirable.

In the context of prediction, the predictive distribution for \(Z_{iu}\) should be understood as being conditional on the event \(\Delta_{iu} = 1\); that is, the prediction is made with the additional information that user \(u\) chose to rate item \(i\). This is the case for nearly all collaborative filtering methods, as correcting for the selection bias necessitates collecting \(Z_{iu}\)'s for which \(\Delta_{iu} = 0\) would occurred naturally; for example, as done by \citet{marlin2009collaborative}, we might assess selection bias by conducting a pilot study which forces users to rate items they would not have normally rated. With the understanding that all methods are predicting ratings conditional on \(\Delta_{iu} = 1\), the results in Section~\ref{sec:data-analysis} show that the multi-rubric model leads to increased predictive performance.

Selection bias should also be taken into account when interpreting the latent rubrics produced by our model. Our model naturally provides a clustering of users into latent classes, which we presented as representing differing standards in user ratings; however, we expect that the model is also detecting differences in selection bias across users. We emphasize that our goal is to identify and account for heterogeneity in rating patterns, and we avoid speculating on whether heterogeneity is caused by different rating standards or selection bias. For example, a user who rates items with only one-star or five-stars might be either (i) using a rubric which results in extreme behavior, with most of the break-points very close together; or (ii) actively choosing to rate items which they feel strongly about. 

\section{Simulation Study}
\label{sec:simulation}

The goal of this simulation is to illustrate that we can accurately learn the existence of multiple rubrics in settings where one would expect it would be difficult to detect them. We consider a situation where the data is generated according to two rubrics that are similar to each other. This allows us to assess the robustness of our model to various ``degrees'' of the multi-rubric assumption. The performance of our multi-rubric model is assessed relative to the single-rubric model, which is the standard assumption made in the ordinal data literature.

We calibrate components of the simulation model towards the \Yelp\ dataset to produce realistic simulated data. Specifically, we set $\eta$ and $\sigma^2_b$ equal to the posterior means obtained from fitting the model to the \Yelp\ dataset in Section~\ref{sec:data-analysis}. We set $\Sigma_\eta = 0.5\Identity$, corresponding to a much stronger spatial effect than what was observed in the data, and for simplicity we removed the latent-factor aspect of the model by fixing $\sigma^2_\beta \equiv 0$. A two-rubric model is used with $\omega_1 = \omega_2 = 0.5$. We also use the same spatial basis functions and observed values of $(i,u)$ as in the \Yelp\ analysis in Section~\ref{sec:data-analysis}.

\begin{figure}
\centering
\includegraphics[width=.8\textwidth]{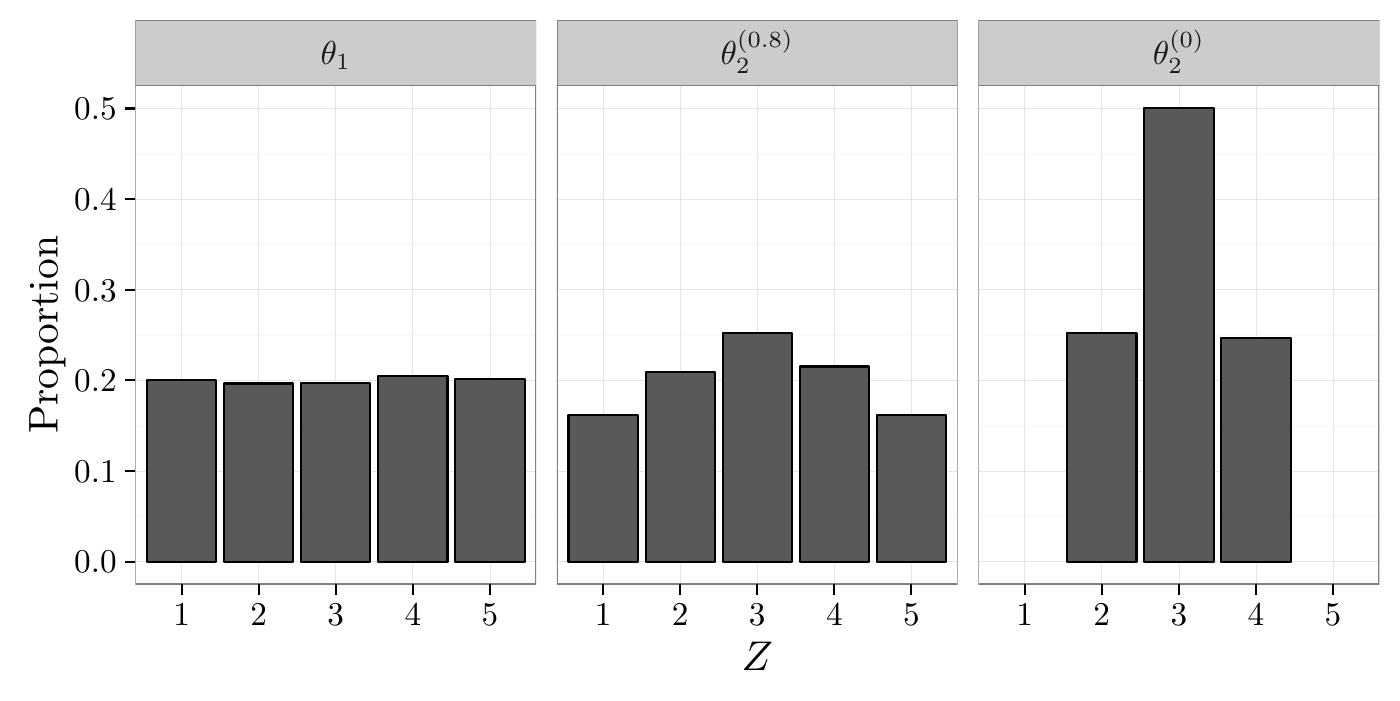}
  \caption{Empirical distribution of the $Z_{iu}$'s in the simulation model, 
  for $\theta_1$, $\theta^{(0.8)}_2$, and $\theta_2^{(0)}$.}
  \label{fig:sim-rubrics}
\end{figure}

We now describe how the two rubrics $\theta_1$ and $\theta_2$ where chosen. First, $\theta_1$ was selected so that $\{Z_{iu} : C_u = 1, i = 1, \ldots, I\}$ was evenly distributed among the five responses. Associated to $\theta_1$ is a probability vector $p_1 = (0.2, 0.2, 0.2, 0.2, 0.2)$. To specify $\theta_2$, we use the same approach with a difference choice of $p$. Let $p_2 = (0, 0.25, 0.5, 0.25, 0)$. Then $\theta_2^{(\tau)}$ is associated to $\tau p_1 + (1 - \tau) p_2$ in the same manner as $\theta_1$ is associated to $p_1$. Here, $\tau$ indexes the similarity of $\theta_1$ and $\theta_2$, and it can be shown that the total variation distance between the empirical distribution of $\{Z_{iu} : C_u = 1\}$ and $\{Z_{iu} : C_u = 2\}$ is $0.8(1 - \tau)$. Thus, values of $\tau$ near $1$ correspond imply that the rubrics are similar, while values of $\tau$ near $0$ imply that they are dissimilar. Figure~\ref{fig:sim-rubrics} presents the distribution of the $Z_{iu}$'s with $C_u = 2$ when $\tau = 0, 0.8$, and $1$.

\begin{figure}[!ht]
\centering
\includegraphics[width = .8\textwidth]{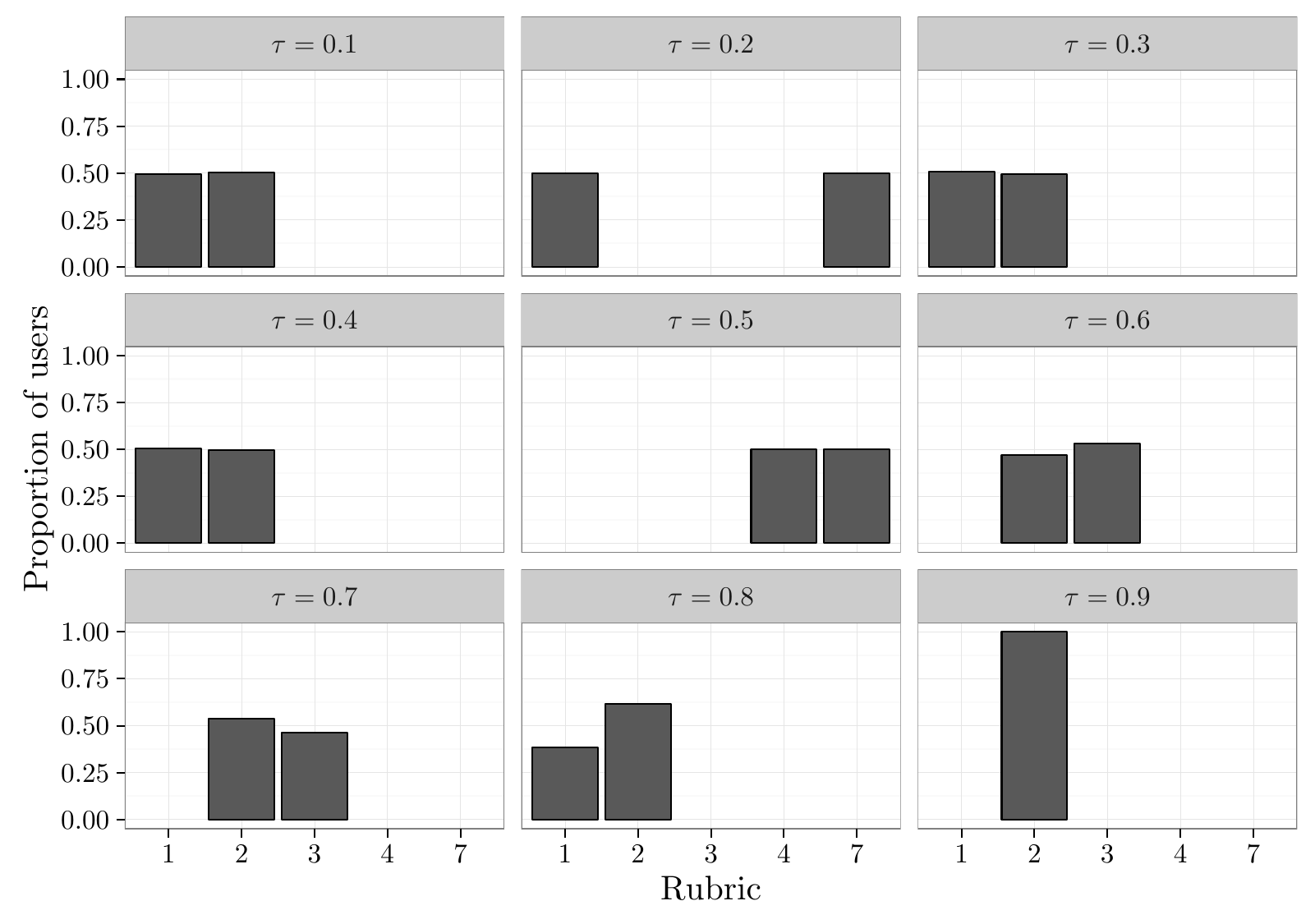}
\includegraphics[width = .8\textwidth]{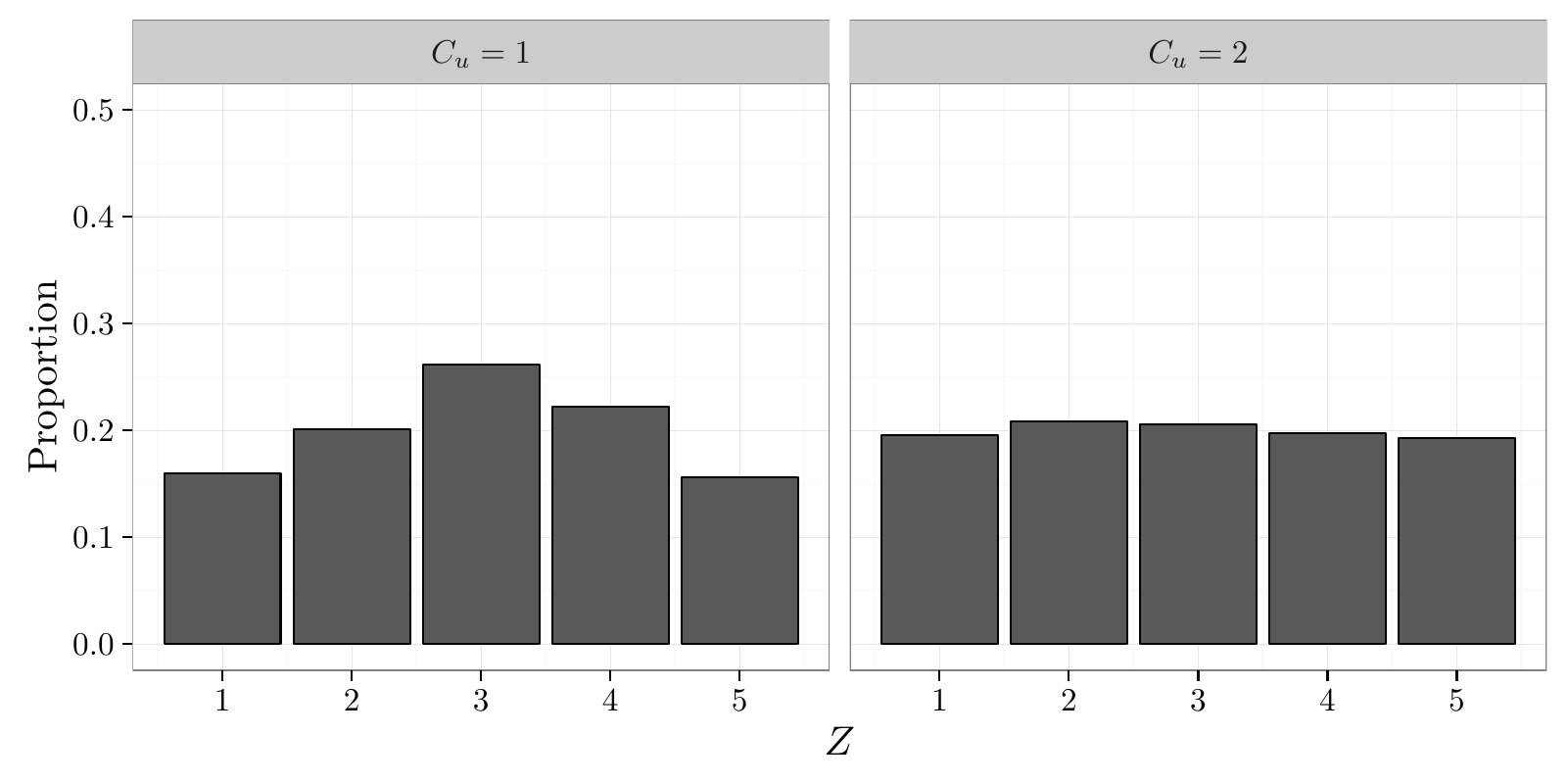}
\caption{
  Top: proportion of individuals assigned to each rubric at the last iteration of the Markov chain. Bottom: The empirical distribution of $Z_{iu}$ for the two rubrics associated to $C_u = 1$ and $C_u = 2$ when \(\tau = 0.8\); compare with the left and middle plots in Figure~\ref{fig:sim-rubrics}.}
\label{fig:apurav-sims}
\end{figure}

We fit a $10$-rubric and single-rubric model for $\tau = 0.0, 0.1, \ldots, 1.0$. Figure~\ref{fig:apurav-sims} displays the proportion of individuals assigned to each rubric for a given value of $\tau$. If the model is accurately recovering the underlying rubric structure, we expect to see a half of the observations assigned to one rubric, and half to another; due to permutation invariance, which of the 10 rubrics is associated to $\theta_1$ and $\theta_2^{(\tau)}$ vary by simulation. Up to $\tau = 0.9$, the model is capable of accurately recovering the existence of two rubrics. We also see that, even at $\tau = 0.8$, the model accurately recovers the empirical distribution of the $Z_{iu}$'s associated to each rubric.

\begin{figure}[!ht]
\centering
\includegraphics[width = 0.75\textwidth]{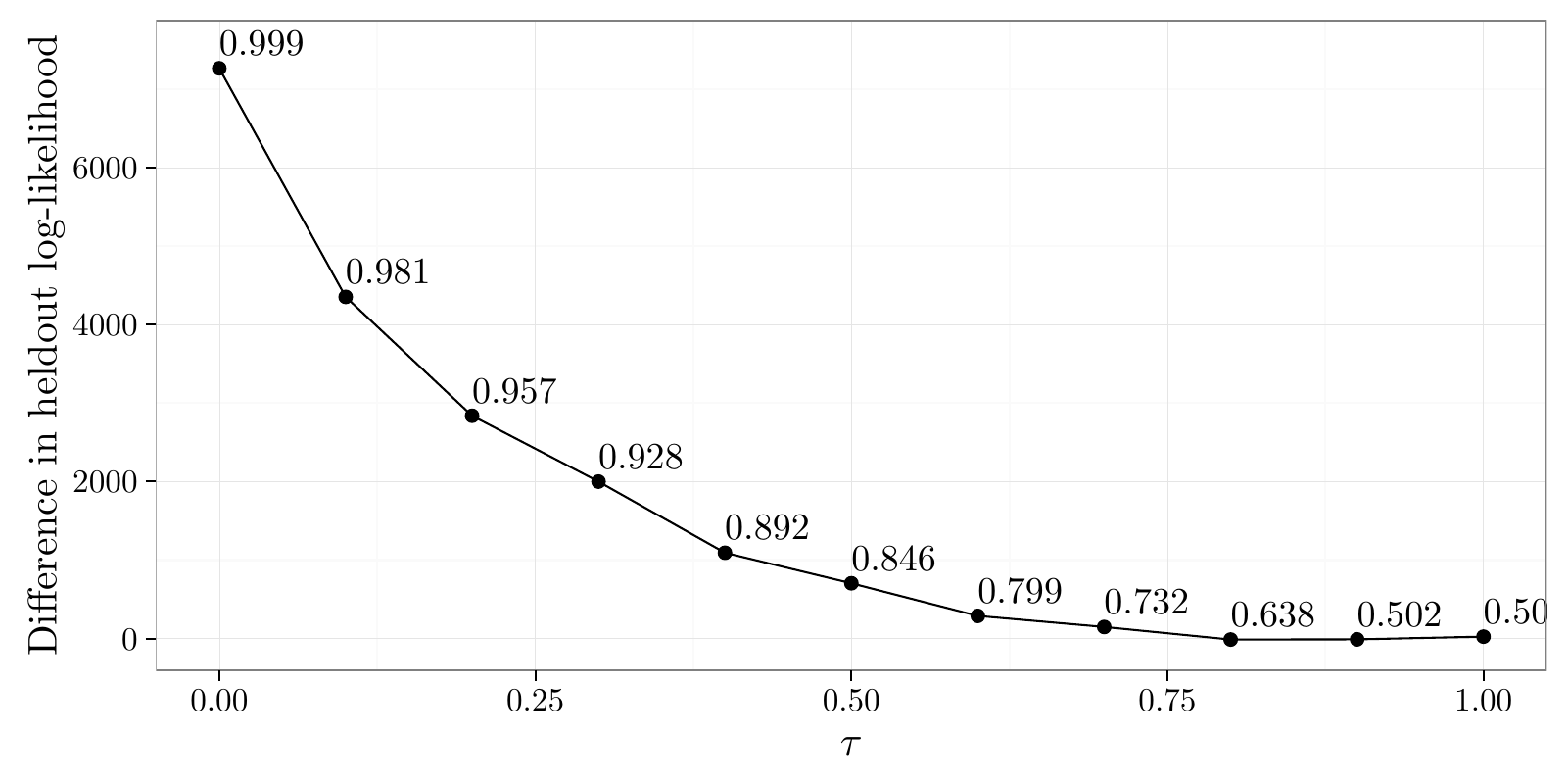}
\caption{
  Difference in $\text{loglik}_{\text{test}}$ for the single-rubric and multi-rubric model obtained in the simulation study, as a function of $\tau$. Above each point, we provide the proportion of users whose most likely rubric assignment matched their true rubric.}
\label{fig:lambda-loss}
\end{figure}

Next, we assess the benefit of using the multi-rubric model to predict missing values. For each value of $\tau$, we fit a single-rubric and multi-rubric model. Using the same train-test split as in the our real data illustration, we compute the log likelihood on the held-out data
\begin{math}
 \text{loglik}_{\text{test}} = \sum_{(i,u) \in \mathcal{S}_{\text{test}}} \log
    \Pr(Z_{iu} \mid D),
\end{math}
which is further discussed in detail in Section~\ref{sec:data-analysis} . Figure~\ref{fig:lambda-loss} shows the difference in held-out log likelihood for the single-rubric and multi-rubric model as a function of $\tau$. Up-to $\tau = 0.8$, there is a meaningful increase in the held-out log-likelihood obtained from using the multi-rubric model. The case where $\tau = 1$ is also particularly interesting, as this implies that the data were generated from the single rubric model. Here the predictive performance of our model at missing values appears to be robust to the case when the multiple rubric assumption is incorrect.

Displayed above each point in Figure~\ref{fig:lambda-loss} is the proportion of observations which are assigned to the correct rubric, where each observation is assigned to their most likely rubric. When the rubrics are far apart the model is capable of accurately assigning observations to rubrics. As the rubrics get closer together, the task of assigning observations to rubrics becomes much more difficult.

This simulation study suggests that the model specified here is able to disentangle the two-rubric structure, even when the rubrics are only subtly different. This leads to clear improvements in predictive performance for small and moderate values for $\tau$. Additionally, when the multi-rubric assumption is negligible, or even incorrect, our model performs as well as the single-rubric model.

\section{Analysis of Yelp data}
\label{sec:data-analysis}

\begin{figure}
  \centering
  \includegraphics[width=.45\textwidth]{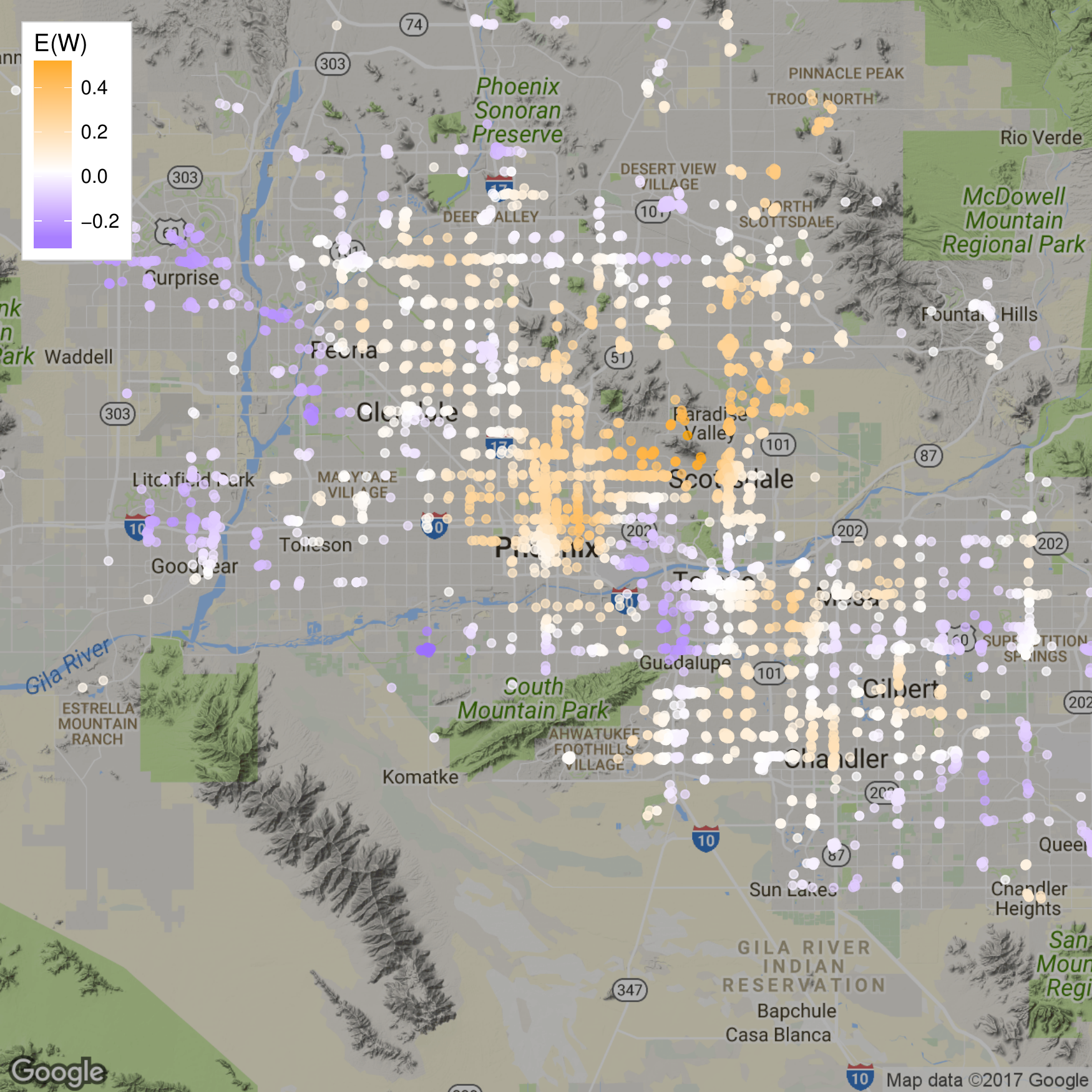}
  \caption{Estimate of the underlying spatial field \(W(s) = \psi(s)\trans\eta\) at each realized restaurant location using its posterior mean. 
  }
  \label{fig:w-process}
\end{figure}

We now apply the multi-rubric model to the \Yelp\ dataset, which is publicly available at \url{https://www.yelp.com/dataset_challenge}. We begin by preprocessing the data to include reviews only between January \(1^{\text{st}}\), 2013 and December \(31^{\text{st}}\) 2016, and restrict attention to restaurants in Phoenix and its surrounding areas. We further narrow the data to include only users who rated at least 10 restaurants; this filtering is done in an attempt to minimize selection bias, as we believe that ``frequent raters'' should be less influenced by selection bias. 

\begin{figure}
  \centering
  \includegraphics[width=1\textwidth]{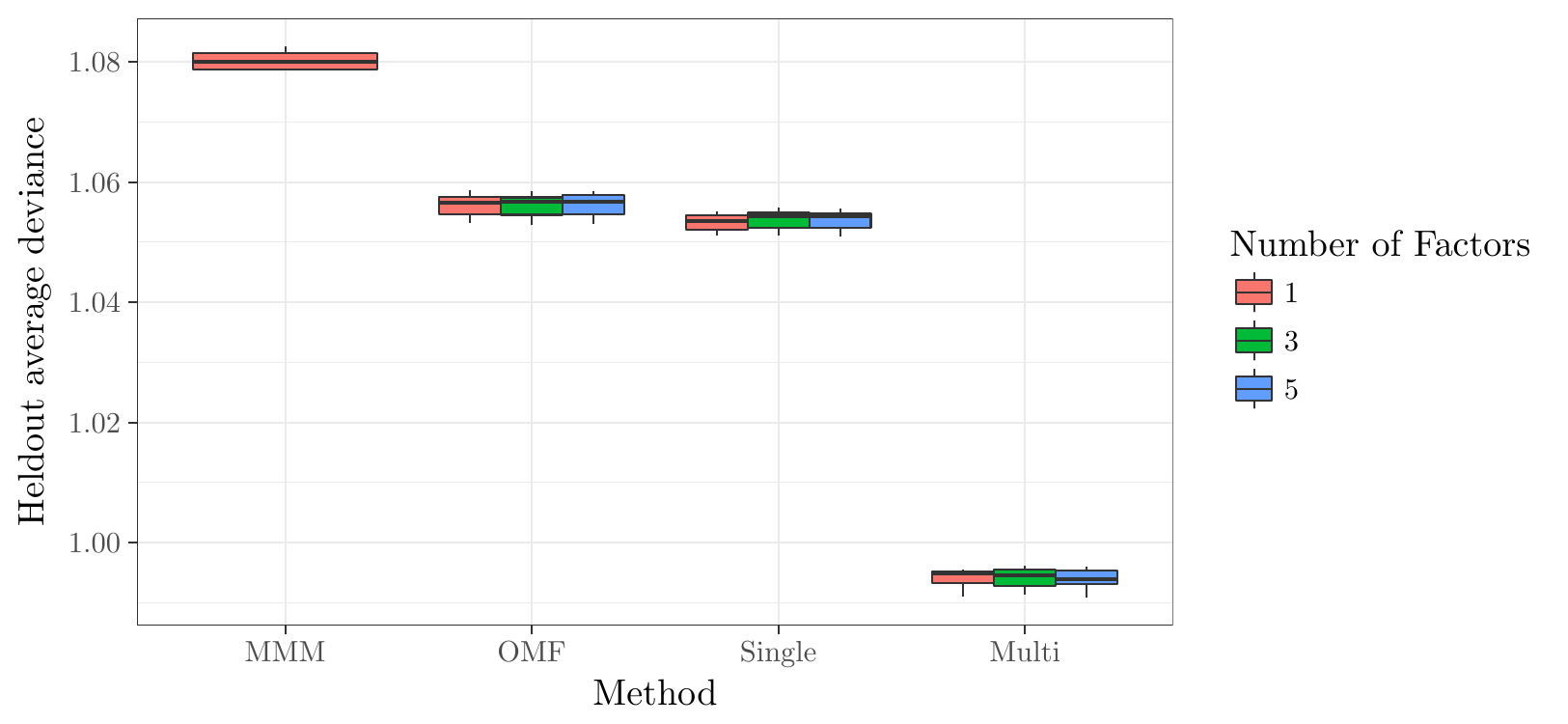}
  \caption{Boxplots of \(-2.0 \cdot \text{loglik}_{\text{test}}\) for the mixture multinomial model (MMM, which does not have latent factors), ordinal matrix factorization (OMF), the single rubric model (Single) and the multi-rubric model (Multi), for 10 splits into training and testing data.}
  \label{fig:heldout-results}
\end{figure}

We first evaluate the performance of the single-rubric and multi-rubric models for various values of the latent factor dimension \(L\). We set \(M = 20\) and induce sparsity in \(\omega\) by setting \(\omega \sim \Dirichlet(1/20, \ldots, 1/20)\). We divide the indices \((i,u) \in \sS\) into a training set \(\sS_{\text{train}}\) and testing set \(\sS_{\text{test}}\) of equal sizes by randomly allocating half of the indices to the training set. We evaluate predictions using a held-out log-likelihood criteria
\begin{align}
  \label{eq:criteria}
  \begin{split}
    \text{loglik}_{\text{test}} &= |\sS_{\text{test}}|^{-1}\sum_{(i,u) \in \sS_{\text{test}}} \log
    \Pr(Z_{iu} \mid \Data), \\&
    \approx |\sS_{\text{test}}|^{-1}\sum_{(i,u) \in \sS_{\text{test}}} \log
    T^{-1} \sum_{t=1}^T \Pr(Z_{iu} \mid C_u^{(t)}, \theta^{(t)}, \mu_{iu}^{(t)}),
  \end{split}
\end{align}
where \(\Data = \{Z_{iu} : (i,u) \in \sS_{\text{train}}\}\), \(\Pr(Z_{iu} \mid \Data)\) denotes the posterior predictive distribution of \(Z_{iu}\), and \(t = 1, \ldots, T\) indexes the approximate draws from the posterior obtained by MCMC. Results for the values \(L = 1, 3\), and 5, over 10 splits into training and test data, are given in Figure~\ref{fig:heldout-results}. We also compare our methodology to ordinal matrix factorization \citep{paquet2012hierarchical} with learned breakpoints and spatial smoothing, and the mixture multinomial model \citep{marlin2009collaborative} with \(10\) mixture components. The multi-rubric model leads to an increase in the held-out data log-likelihood \eqref{eq:criteria} of roughly \(5\%\) over ordinal matrix factorization and \(8\%\) over the mixture multinomial model. Additionally, we note that the holdout log-likelihood was very stable over replications. The single-rubric model is essentially equivalent to ordinal matrix factorization.

The dimension of the latent factors \(\alpha_u\) and \(\beta_i\) has little effect on the quality of the model. We attribute this to the fact that \(|\sU_i|\) and \(\sI_u|\) are typically small, making it difficult for the model to recover the latent factors. On other datasets where this is not the case, such as the Netflix challenge dataset, latent-factor models represent the state of the art and are likely essential for the multi-rubric model. In the supplementary material we show in simulation experiments that the \(\alpha_u\)'s, \(\beta_i\)'s, and \(L\) are identified.

Figure~\ref{fig:w-process} displays the learned spatial field \(\widehat W(s) = \psi(s)\trans\widehat\eta\) where \(\widehat \eta\) is the posterior mean of \(\eta\).
The results suggest that the downtown Phoenix business district and the area surrounding the affluent Paradise Valley possesses a higher concentration of highly-rated restaurants than the rest of the Phoenix area. More sparsely populated areas such as such as Litchfield Park, or areas with lower income such as Guadalupe, seem to have fewer highly-rated restaurants.

\begin{figure}[!t]
  \centering
  \includegraphics[width = .9\textwidth]{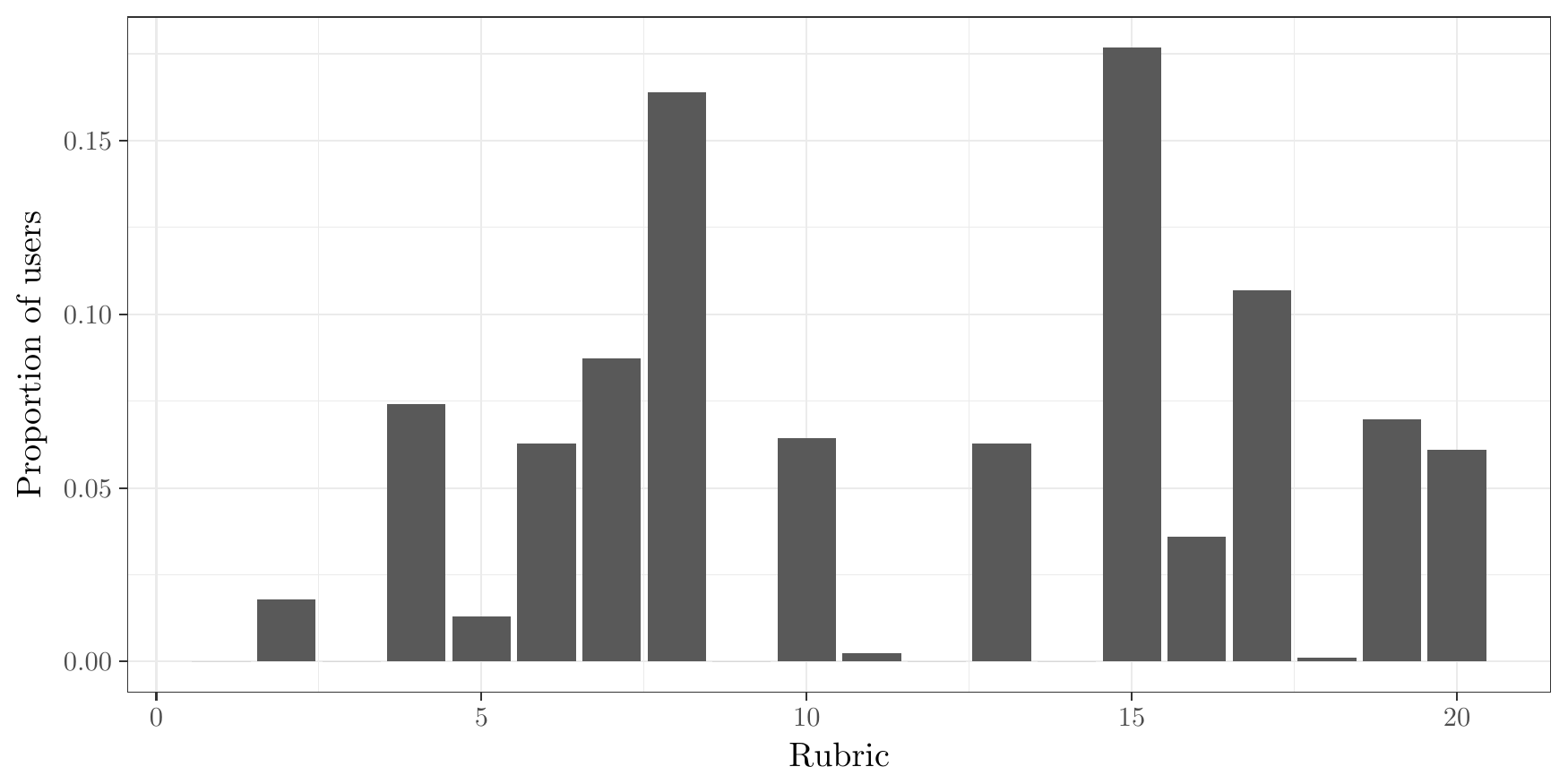}
  \includegraphics[width=.9\textwidth]{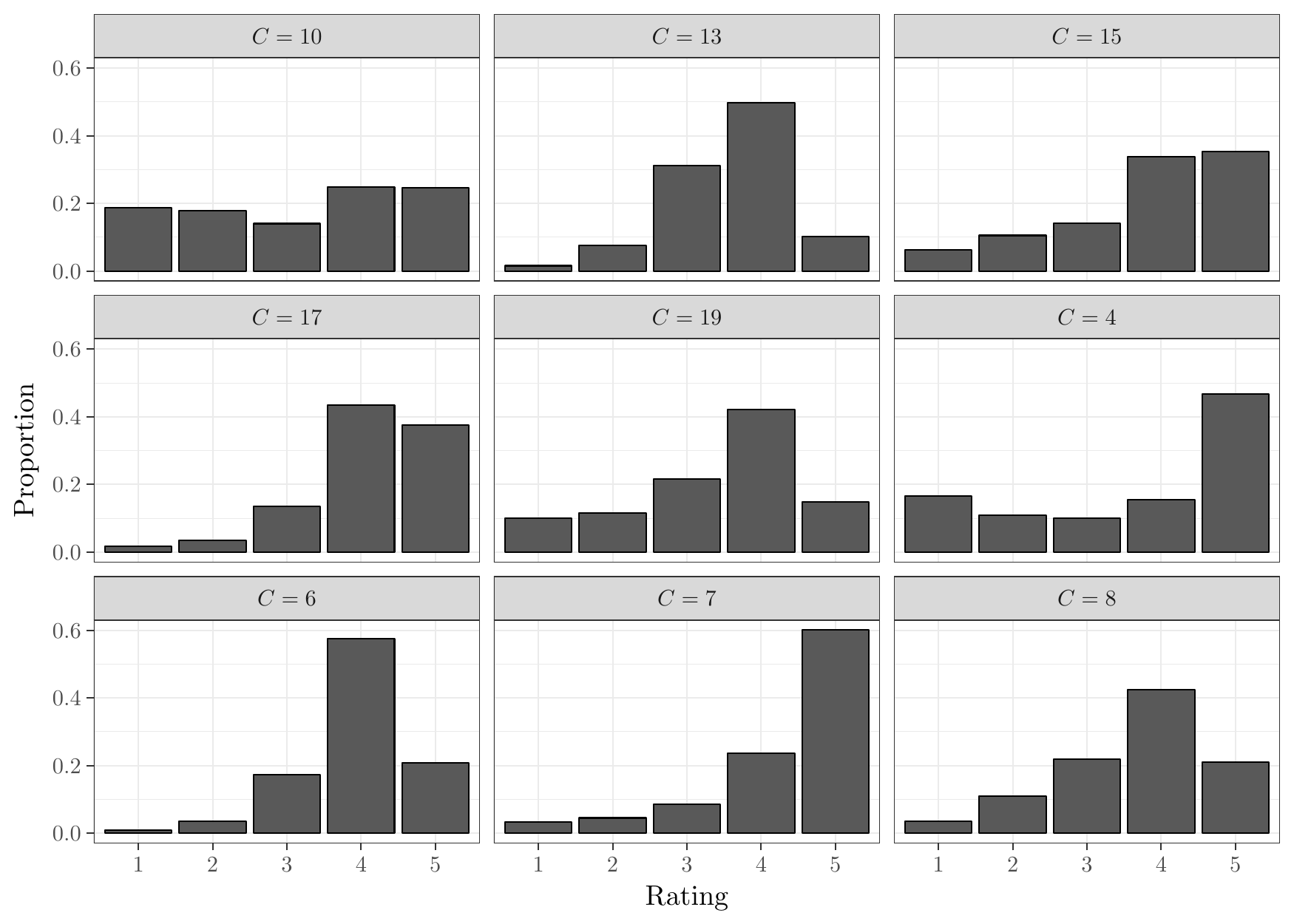}
  \caption{
    Top: bar chart giving the number of users assigned to each rubric, where users are assigned to rubrics by minimizing Binder's loss function. Bottom: bar charts giving the proportions of the observed ratings \(Z_{iu}\) for each item-user pair for the top 9 most common rubrics.}
  \label{fig:rubric-props}
\end{figure}

We now examine the individual rubrics. First, we obtain a clustering of users into their rubrics by minimizing Binder's loss function \citep{binder1978bayesian} $L(\bm c) = \sum_{u,u'} |\delta_{c_u, c_{u'}} - \Pi_{u,u'}$, where $\delta_{ij} = I(i = j)$ is the Kronecker delta, $\bm c = (c_1, \ldots, c_U)$ is an assignment of users to rubrics, and $\Pi_{u,u'}$ is the posterior probability of $C_u = C_{u'}$. See \citet{fritsch2009improved} for additional approaches to clustering objects using samples of the $C_u$'s.

The multi-rubric model produces interesting effects on the overall estimate of restaurant quality. Consider the rubric corresponding to \(m = 7\) in Figure~\ref{fig:rubric-props}. Users assigned to this rubric give the majority of restaurants a rating of five stars. As a result, a rating of 5 stars for the \(m = 7\) rubric is less valuable to a restaurant than a rating of 5 stars from a user with the \(m = 6\) rubric. Similarly, a rating of \(3\) stars from the \(m = 7\) rubric is more damaging to the estimate of a restaurant's quality than a rating of \(3\) stars from the \(m = 6\) rubric.

\begin{figure}
  \centering
  \includegraphics[width=.8\textwidth]{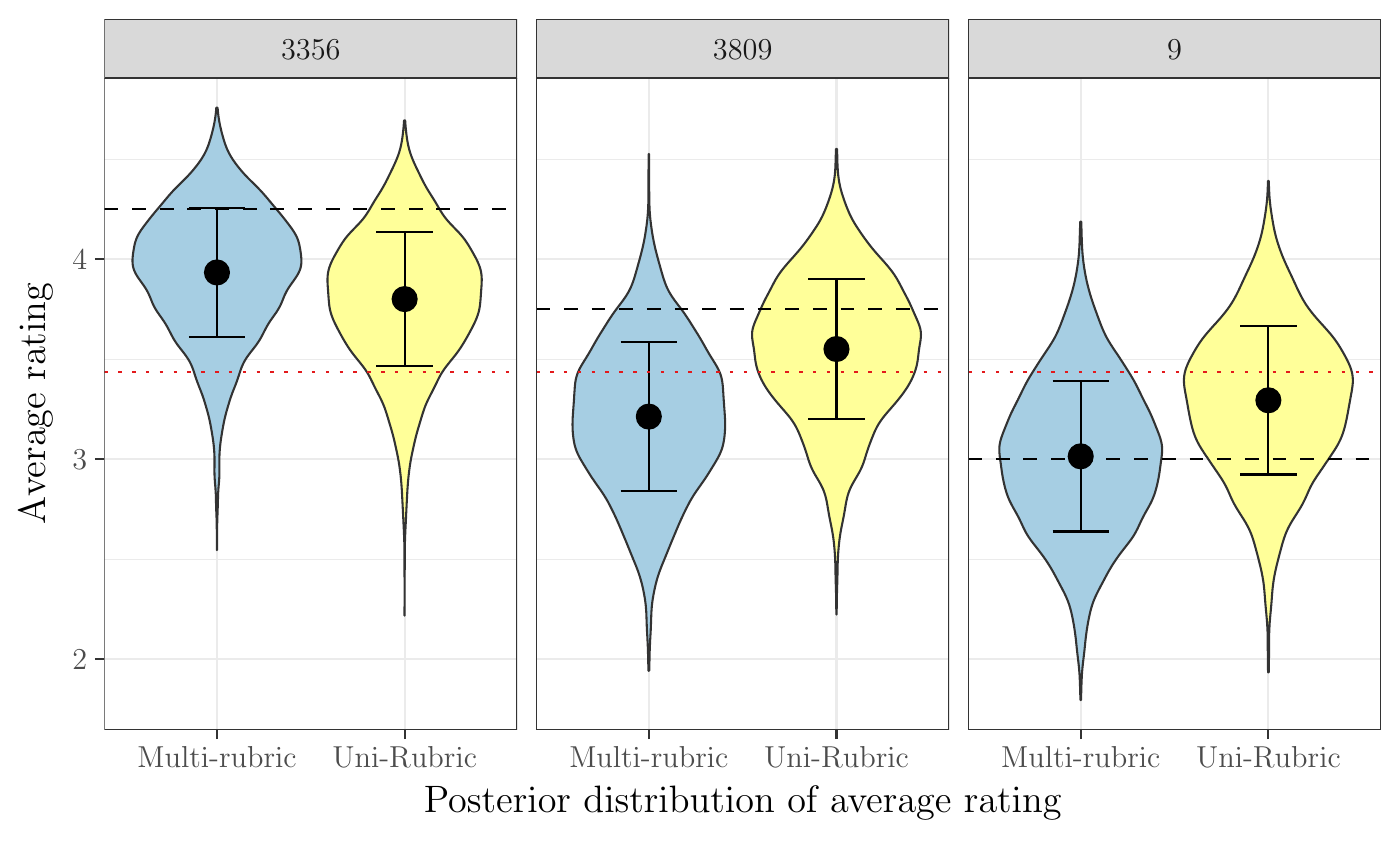}
  \caption{Posterior density of \(\lambda_i\) for \(i \in \{9, 3356, 3809\}\).
    The dashed line is the empirical average rating of item \(i\); the dotted
    line is the overall average of all ratings. Error bars are centered at the
    posterior mean with a radius of one standard deviation.}
  \label{fig:rating-posterior}
\end{figure}

For restaurants with a large number of reviews, the effect mentioned above is negligible, as the restaurants typically have a good mix of users from different rubrics. The effect on restaurants with a small number of reviews, however, can be much more pronounced. To illustrate this effect, Figure~\ref{fig:rating-posterior} displays the posterior distribution of the quantity \(\lambda_i\) defined in \eqref{eq:lambda} for the restaurants with \(i \in \{ 3356, 3809, 9\}\). Each of these businesses has \(4\) reviews total, with empirically averaged ratings of 4.25, 3.75, and 3 stars. For \(i = 3809\) and \(i = 9\), the users are predominantly from the rubric with \(m = 7\); as a consequence, the fact that these restaurants do not have an average rating closer to five stars is quite damaging to the estimate of the restaurant quality. In the case of \(i = 3809\), the effect is strong enough that what was ostensibly an above-average restaurant is actually estimated to be below average by the multi-rubric model. Conversely, item \(i = 3356\) has ratings of \(4, 5, 5\), and \(3\) stars, but one of the \(5\)-star ratings comes from a user assigned to the rubric \(m = 2\) which gave a \(5\)-star rating to only 8\% of businesses. As a result, the \(5\)-star ratings are weighted more heavily than they would otherwise be, causing the distribution of \(\lambda_i\) to be shifted slightly upwards.  

\begin{figure}
  \centering
  \includegraphics[width=.8\textwidth]{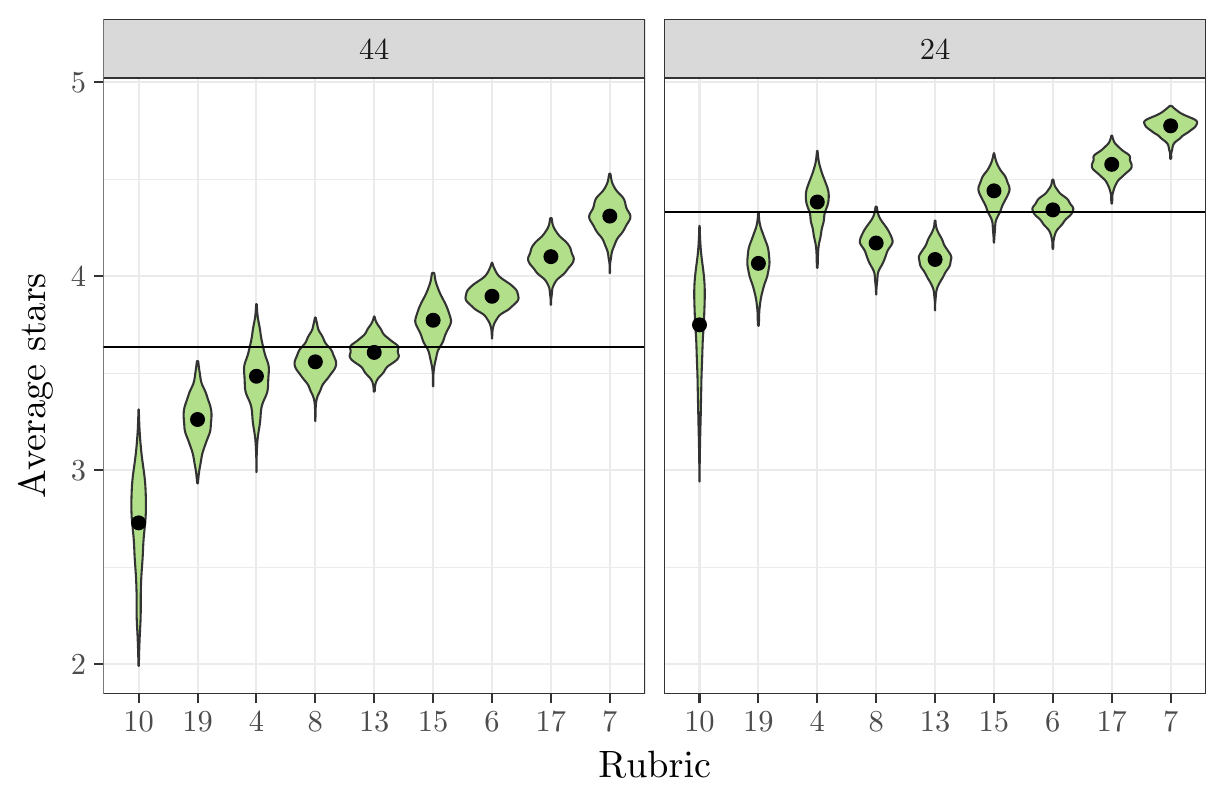}
  \caption{Posterior density of \(\lambda_{im}\) for \(i = 44, 24\). Horizontal
    lines display the empirical average rating for each item. Rubrics are
    organized by the average rating assigned to \(i = 44\) for visualization
    purposes.}
    \label{fig:rubric-specific}
\end{figure}

Lastly, we consider rescaling the average ratings according to a specific rubric. This may of interest, for example, if one is interested in standardizing the ratings to match a rubric which evenly disperses ratings evenly across the possible stars. To do this, we examine the rubric-adjusted average ratings \(\lambda_{im}\) given by \eqref{eq:lambda-k}. Figure~\ref{fig:rubric-specific} displays the posterior density of \(\lambda_{im}\) for \(i = 24\) and \(i = 44\), for the \(9\) most common rubrics. These two restaurants have over 100 reviews, and so the overall quality can be estimated accurately. We see some expected features; for example, the quality of each restaurant has been adjusted downwards for users of the \(m = 10\) rubric, and upwards for the \(m = 7\) rubric. The multi-rubric model allows for more nuanced behavior of the adjusted ratings than simple upward/downward shifts. For example, for the mediocre item \(i = 44\), we see that little adjustment is made for the \(m = 13\) rubric, while for the high-quality item \(i = 24\) a substantial downward adjustment is made. This occurs because the model interprets the users with \(m = 13\) as requiring a relatively large amount of utility to rate an item 5 stars, so that a downward adjustment is made for the high-quality item; on the other hand, users with \(m = 13\) tend to rate things near a \(3.5\), so little adjustment needs to be made for the mediocre item.

\section{Discussion}
\label{sec:discussion}

In this paper we introduced the multi-rubric model for the analysis of rating data and applied it to public data from the website \Yelp. We found that the multi-rubric model yields improved predictions and induces sophisticated shrinkage effects on the estimated quality of the items. We also showed how the model developed can be used to partition the users into interpretable latent classes.

There are several areas exciting areas for future work. First, while Markov chain Monte Carlo works well for the \Yelp\ dataset (it took 90 minutes to fit the model of Section~\ref{sec:data-analysis}), it would be desirable to develop a more scalable procedure, such as stochastic variational inference \citep{hoffman2013stochastic}. Second, the model described here features limited modeling of the users. Information regarding which items the users have rated has been shown in other settings to improve predictive performance; temporal heterogeneity may also be present in users.

The latent class model described here can also be extended to allow for more flexible models for the \(\alpha_u\)'s and \(\beta_i\)'s. For example, a referee pointed out the possibility of inferring about how controversial an item is across latent classes, which could be accomplished naturally by using a mixture model for the \(\alpha_u\)'s.


A fruitful area for future research is the development of methodology for when MAR fails. One possibility for future work is to extend the model to also model the missing data indicators \(\Delta_{iu}\). This is complicated by the fact that, while \(\{Y_{iu} : 1 \le i \le I, 1 \le u \le U\}\) is not completely observed, \(\{\Delta_{iu} : 1 \le i \le I, 1 \le u \le U\}\) is. As a result, the data becomes much larger when modeling the \(\Delta_{iu}\)'s.

\ifblinded
\else
\section*{Acknowledgements}
The authors thank Eric Chicken for helpful discussions. This work was partially supported by the Office of the Secretary of Defense under research program \#SOT-FSU-FATs-06 and NSF grants NSF-SES \#1132031 and NSF-DMS \#1712870.
\fi

\appendix

\section{Markov chain Monte Carlo algorithm}

Before describing the algorithm, we define several quantities. First, define
\begin{alignat*}{3}
  R_{iu}^{(\alpha)} &= Y_{iu} - \mu_{iu} + \alpha_i\trans\beta_u, &&\qquad&
  R_{iu}^{(b)} &= Y_{iu} - \mu_{iu} + b_i, \\
  R_{iu}^{(\gamma)} &= Y_{iu} - \mu_{iu} + x_i\trans\gamma, &&\qquad&
  R_{iu}^{(\eta)} &= Y_{iu} - \mu_{iu} + \psi(s_i)\trans\eta.
\end{alignat*}
Let
\begin{math}
  \bm R^{(\alpha)}_u = \vec(R_{iu}^{(\alpha)} : i \in \sI_i),
  \bm R^{(\beta)} = \vec(R_{iu}^{(\alpha)} : u \in \sU_i), 
  \bm R^{(b)}_i = \vec(R_{iu}^{(b)} : u \in \sU_i), 
  \bm R^{(\gamma)} = \vec(R_{iu}^{(\gamma)} : (i,u) \in \sS), 
  \text{ and }
  \bm R^{(\eta)}_i = \vec(R_{iu}^{(\eta)} : (i,u) \in \sS).
\end{math}
Then we can write
\begin{alignat*}{3}
  \bm R^{(\beta)}_i  &= \bm A_i \beta_i + \bm \epsilon_i, &&\qquad&
  \bm R^{(b)}_i  &= \ones b_i + \bm \epsilon_i, \\
  \bm R^{(\alpha)}_u &= \bm B_u \alpha_i + \bm \epsilon_i, &&\qquad&
  \bm R^{(\gamma)}_u &= \bm X \gamma + \bm \epsilon, \\
  \bm R^{(\eta)}_u &= \bm \Psi \eta + \bm \epsilon,
\end{alignat*}
where \(\bm A_i\) has rows composed of \(\alpha\)'s associated to users who rated item \(i\), \(\bm B_u\) has rows composed of \(\beta\)'s associated to items which were rated by user \(u\), and \(\bm X\) and \(\bm \Psi\) are design matrices associated to the covariates and basis functions respectively. Holding the other parameters fixed, each term above on the left-hand-side is sufficient for its associated parameter on the right-hand-side.

A data augmentation strategy similar to the one proposed by \citet{albert1997bayesian} is employed. The updates for the parameters \(\eta, \alpha, \beta\), and \(\gamma\) use a back-fitting strategy based on the \(\bm R\)'s above. The Markov chain operates on the state space \( ( C, Y, \theta, b, \alpha, \beta, \gamma, \eta, \omega, \sigma_\alpha, \sigma_\beta, \sigma_\eta )\). We perform the following updates for each iteration of the sampling algorithm, where each step is understood to be done for each relevant \(u\) and \(i\).
\begin{enumerate}
\item Draw \(C_u \sim \Categorical(\widehat \omega_{u1}, \ldots, \widehat
  \omega_{uM})\) where \(\widehat \omega_{um}\) is proportional to \(\omega_m
  \prod_{i \in \mathcal U_u} [\Phi(\theta_{Z_{iu}}^{(m)} - \mu_{iu}) -
  \Phi(\theta_{Z_{iu}-1}^{(m)} - \mu_{iu})]\)
\item Draw \(Y_{iu} \sim \TruncatedNormal(\mu_{iu}, 1, \theta^{(C_u)}_{k-1},
  \theta^{(C_u)}_k)\), for \((i,u) \in \mathcal S\).
\item Draw \(\alpha_u \sim \Normal(\widehat \alpha_u, \widehat
  \Sigma_{\alpha_u})\) where
  \begin{math}
    \widehat \Sigma_{\alpha_u}
    =
    (\bm B_u\trans \bm B_u + \sigma_\alpha^2 I)^{-1}
  \end{math}
  and \(\widehat \alpha_u = \widehat \Sigma_{\alpha_u} \bm B_u\trans \bm R^{(\alpha)}_u\).
\item Draw \(\beta_i \sim \Normal(\widehat \beta_i, \widehat \Sigma_{\beta_i})\)
  where
  \begin{math}
    \widehat \Sigma_{\beta_i}
    =
    (\bm A_i\trans \bm A_i + \sigma^2_\beta I)^{-1}
  \end{math}
  and \(\widehat \beta_i = \widehat \Sigma_{\beta_i} \bm A_i\trans \bm R^{(\beta)}_i\).
\item Draw \(\gamma \sim \Normal(\widehat \gamma, \widehat \Sigma_\gamma)\)
  where \(\widehat \Sigma_\gamma = (\bm X\trans \bm X)^{-1}\) and \(\widehat
  \gamma = \widehat \Sigma_\gamma \bm X\trans \bm R^{(\gamma)}\).
\item Draw \(b_i \sim \Normal(\widehat b_i, \widehat \sigma^2_{b_i})\) where
  \begin{math}
    \widehat \sigma^2_{b_i} = (\sigma^{-2}_b + |\sU_i|)^{-1},
  \end{math}
  and \(\widehat b_i = \widehat \sigma^2_{b_i} \sum_{u \in \sU_i} R_{iu}^{(b)}\).
\item Draw \(\eta \sim \Normal(\widehat \eta, \widehat \Sigma_\eta)\) where
  \begin{math}
    \widehat \Sigma_\eta
    =
    (\bm\Psi\trans\bm\Psi + \Sigma^{-1}_\eta)^{-1}
  \end{math}
  and \(\widehat \eta = \widehat \Sigma_\eta \bm \Psi\trans \bm R^{(\eta)}\).
\item Draw \(\omega \sim \Dirichlet(\widehat a_1, \ldots, \widehat
  a_M)\) where \(\widehat a_m = a + \sum_{u : C_u = m} 1\).
\item Make an update to \(\sigma^2_b\) which leaves
  \begin{math}
    \Gam(\sigma^2_b \mid 0.5, 0.5) \prod_{i=1}^I \Normal(b_i \mid 0, \sigma_b^2)
  \end{math}
  invariant.
\item Make an update to \(\sigma^2_\beta\) which leaves
  \begin{math}
    \Gam(\sigma^2_{\beta} \mid 0.5, 0.5) \prod_{i=1}^I \prod_{\ell = 1}^L
    \Normal(\beta_{i\ell} \mid 0, \sigma_\beta^2)
  \end{math}
  invariant.
\item Make an update to \(\sigma^2_\eta\) which leaves
  \begin{math}
    \Gam(\sigma^2_{\eta} \mid 0.5, 0.5) \prod_{j=1}^r 
    \Normal(\eta_j \mid 0, \sigma_\eta^2)
  \end{math}
  invariant.
\item Make an update to \(\theta^{(m)}\) which leaves
  \begin{math}
    \Normal(\theta^{(m)} \mid \zeros, \sigma_\theta^2 \Identity) I(\theta_1^{(m)} < \ldots, < \theta_{K-1}^{(m)})
    \cdot \prod_{u : C_u = m}
    \prod_{i \in \sI_u}
    \log [\Phi(\theta_{Z_{iu}}^{(m)} - \mu_{iu}) - \Phi(\theta^{(m)}_{Z_{iu}-1} - \mu_{iu})],
  \end{math}
  invariant.
\end{enumerate}
In our illustrations, we use slice sampling \citep{slicesampling} to do updates 9--11. The chain is initialized by simulation from the prior with \(a = 1\). The only non-trivial step is constructing an update for the \(\theta^{(m)}\)'s. We use a modification of the approach outlined in \citet{albert1997bayesian}, which uses a Laplace approximation to construct a proposal for the full-conditional of the parameters \(\delta^{(m)}_1 = \theta^{(m)}_1\) and \(\delta^{(m)}_k = \log(\theta^{(m)}_k - \theta^{(m)}_{k-1})\) for \(k = 2,\ldots,K-1\). To alleviate computational burden, the proposal is updated every \(50^{\text{th}}\) iteration.

\bibliographystyle{apalike}
\bibliography{myref.bib,mybib.bib}

\clearpage 

\title{Supplementary Material}
\author{Antonio R. Linero, Jonathan R. Bradley, Apruva S. Desai}

\maketitle

\pagestyle{fancy}
\fancyhf{}
\rhead{\bfseries\thepage}
\lhead{\bfseries NOT FOR PUBLICATION SUPPLEMENTARY MATERIAL}
\setcounter{equation}{0}
\setcounter{page}{1}
\setcounter{table}{1}
\setcounter{section}{0}
\renewcommand{\theequation}{S.\arabic{equation}}
\renewcommand{\thesection}{S.\arabic{section}}
\renewcommand{\thesubsection}{S.\arabic{section}.\arabic{subsection}}
\renewcommand{\thepage}{S.\arabic{page}}
\renewcommand{\thetable}{S.\arabic{table}}

\thispagestyle{fancy}

\section{Identifiability}

We conduct a brief simulation experiment to illustrate that model proposed in Section~\ref{sec:description} is capable of (i) identifying the correct number of latent factors \(L\) and (ii) capable of accruing evidence about the individual \(\alpha_u\)'s and \(\beta_i\)'s. We simulate from the data model, response model, random effect model, spatial process model, and parameter model described in Section~\ref{sec:description} with the dimension of the latent factor set to \(L = 4\). For simplicity, we omit the item-specific covariates given by \(x_i\). For the random effects model we set \(\sigma_\alpha = 2, \sigma_\beta = 5\), and \(\sigma_b = 3\). We set \(\eta \sim \Normal(0, 3\Identity)\) and \(r = 20\), with the basis functions \(\psi_j(s)\) given by Gaussian radial basis functions with knots at placed uniformly at random throughout the spatial domain. For the parameter model, we used \(M = 3\) rubrics with \(\omega = (1/3, 1/3, 1/3)\) and selected \(\theta^{(m)}\) in the manner described in the simulation of Section~\ref{sec:simulation}. We set \(u = 200\) and \(i = 200\), and select user/item pairs for inclusion by sampling uniformly at random, with a total of 3981 ratings. 

After simulating this data, we fit the multi-rubric model using the correct \(\Psi\) using the default prior described in Section~\ref{sec:description} with the correct choice of basis functions \(\psi_j(s)\), with \(M = 10\) and \(\omega \sim \Dirichlet(1/10, \ldots, 1/10)\).

\begin{figure}
  \centering
  \includegraphics[width = .49\textwidth]{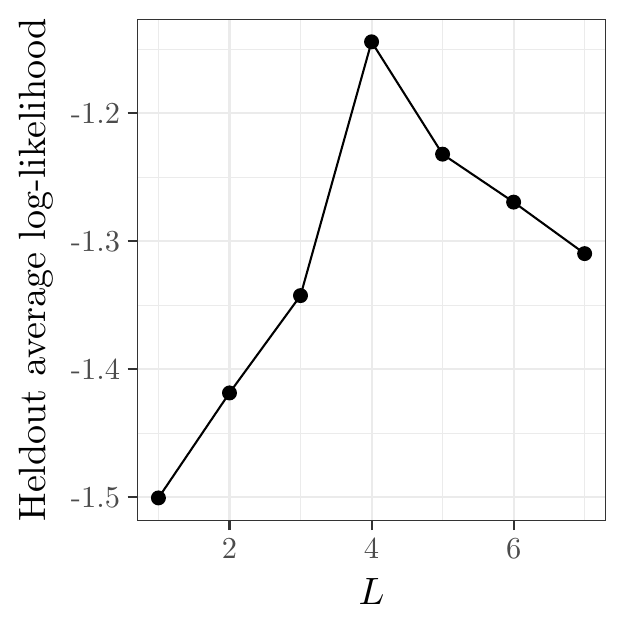}
  \caption{Held-out log likelihood for different values of \(L\).}
  \label{fig:loss}
\end{figure}

We first assess whether the model is capable of recovering the true number of latent factors. We fit the model for \(L \in \{1, \ldots, 7\}\) and evaluated each value of \(L\) by held-out log-likelihood criteria \eqref{eq:criteria} after splitting the data randomly into training and testing sets. Results are given in Figure~\ref{fig:loss}. We see that the model with the highest held-out log-likelihood corresponds to \(L = 4\), the correct number of latent factors.

Next, we assess whether the model is capable of learning the individual \(\alpha_u\)'s and \(\beta_i\)'s. First, we note that for any orthonormal matrix \(O\) with \(O\trans O = \Identity\) we have
\begin{align*}
  \alpha_u\trans\beta_i = (O\alpha_u)\trans (O\beta_u).
\end{align*}
Moreover, $\alpha_u$ and $O\alpha_u$ are equal in distribution (as are $\beta_i$ and $O\beta_i$), so the prior does not provide any additional identification. Consequently, we can only expect to identify \(\alpha_u\) and \(\beta_i\) up-to an arbitrary rotation. While it is possible to impose constraints on the \(\alpha_u\) and \(\beta_i\) --- say, by fixing \(\alpha_1, \ldots, \alpha_L\) to know values --- this is undesirable because it breaks the symmetry of the prior. In view of this, it is standard in the recommender systems literature to not invoke any constraints on the prior \citepNew{salakhutdinov2007probabilistic}.

\begin{figure}
  \centering
  \includegraphics[width = 1\textwidth]{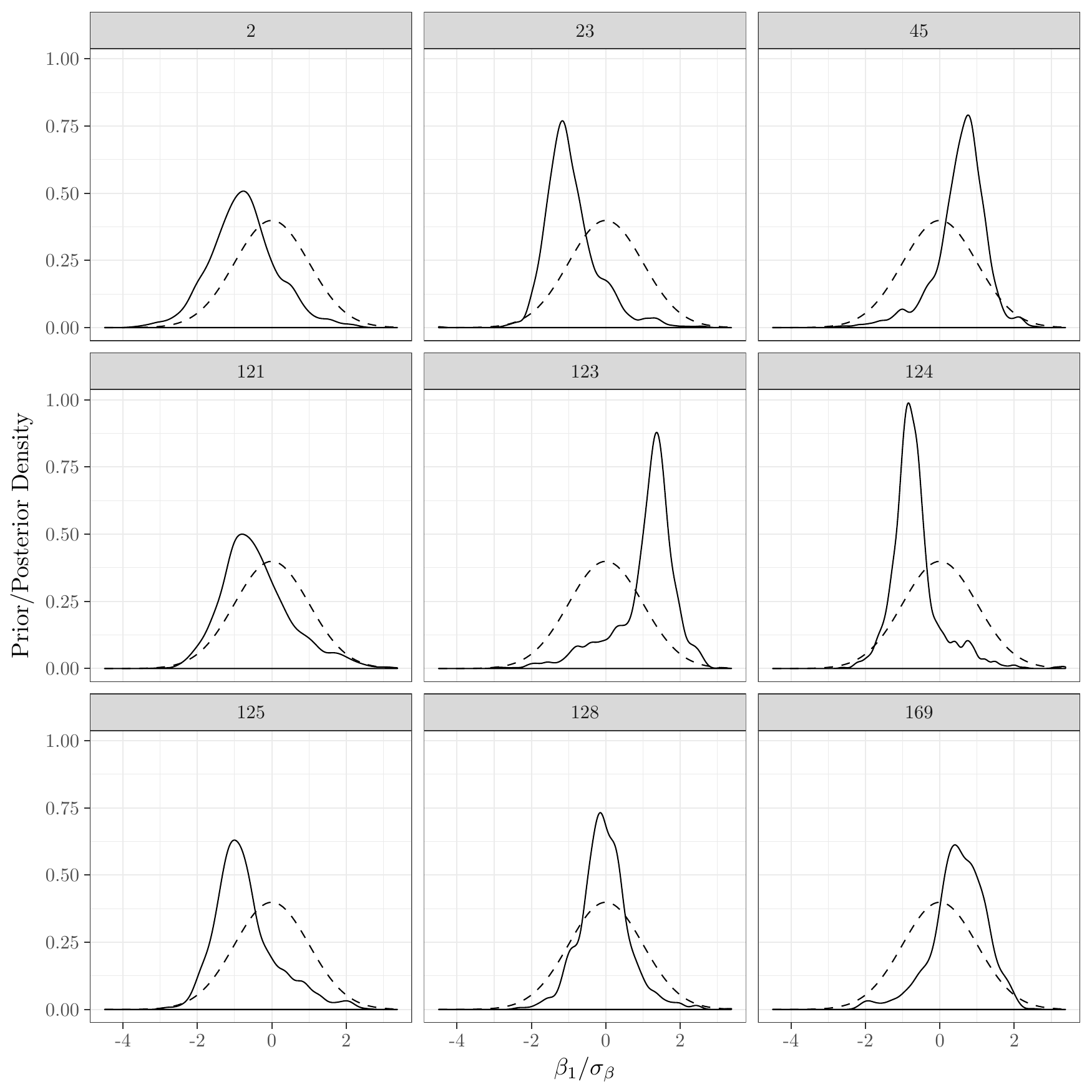}
  \caption{Posterior density estimate of \(\zeta_i = \beta_{i1} / \sigma_\beta\) for randomly selected items (solid) and the prior density \(\zeta_i \sim \Normal(0,1)\) (dashed).}
  \label{fig:learning}
\end{figure}

With these points in mind, Figure~\ref{fig:learning} displays the prior distribution of \(\zeta_i = \beta_{i1} / \sigma_{\beta}\) along with the posterior distribution of \(\zeta_i\)'s for 9 randomly selected items. We see that, while the overall distribution of the \(\zeta_i\)'s is in agreement with the prior when considered as a group, for the individual \(\zeta_i\)'s the prior and posterior differ considerably. This indicates that the model is capable of detecting differences in the \(\beta_i\)'s across items.

\bibliographystyleNew{apalike}
\bibliographyNew{mybib.bib}

\end{document}